\documentclass[journal,12pt,onecolumn,draftcls]{IEEEtran}

\usepackage{cite}
\usepackage{graphics}
\usepackage{url}
\usepackage{amsmath}
\usepackage{amssymb}
\usepackage{amsthm}
\usepackage{array}
\usepackage{epsfig}

\newcommand{\comment}[1]  {  }

\def\BE{\begin{equation}}
\def\EE{\end{equation}}
\def\BEA{\begin{eqnarray}}
\def\EEA{\end{eqnarray}}

\newcommand{\fd}[2]{\frac{d #1}{d #2}}
\newcommand{\pd}[2]{\frac{\partial #1}{\partial #2}}

\newtheorem*{GSV}{GSV Theorem}
\newtheorem*{LOT}{Laws of thermodynamics}
\newtheorem*{first}{$\mathbf{1^{st}}$ Law}
\newtheorem*{second}{$\mathbf{2^{nd}}$ Law}
\newtheorem*{second_G}{Generalized $\mathbf{2^{nd}}$ Law}
\newtheorem*{third}{$\mathbf{3^{rd}}$ Law}

\newcommand{\snr}{\textsf{snr}}

\newcommand{\mmse}{\textsf{mmse}}
\newcommand{\ie}{\textsl{i.e.}}
\newcommand{\eg}{\textsl{e.g.}}
\newcommand\etal{{\textsl{et al.\,}}}
\newcommand{\dbar}{d\mkern-6mu\mathchar'26}

\def\thefootnote{\fnsymbol{footnote}}

\begin{document}

\title{Shannon Meets Carnot:\\Mutual Information Via Thermodynamics}

\author{Ori~Shental\thanks{O. Shental is with the Center for Magnetic Recording Research (CMRR), University of California, San Diego (UCSD), 9500 Gilman Drive, La Jolla, CA 92093, USA (e-mail: oshental@ucsd.edu).} and
        Ido~Kanter\thanks{I. Kanter is with the Minerva Center and Department of Physics, Bar-Ilan University, Ramat-Gan 52900, Israel (e-mail: kanter@mail.biu.ac.il).}
}

\maketitle



\date{}
\maketitle

\begin{abstract} In this contribution, the Gaussian channel is represented as an equivalent thermal system allowing to express its input-output mutual information in terms of thermodynamic quantities. This thermodynamic description of the mutual information is based upon a generalization of the $2^{nd}$ thermodynamic law and provides an alternative proof to the Guo-Shamai-Verd\'{u} theorem, giving an intriguing connection between this remarkable theorem and the most fundamental laws of nature - the laws of thermodynamics.
\end{abstract}

\newline

\renewcommand{\thefootnote}{\arabic{footnote}}
\setcounter{footnote}{0}

\section{Introduction}
The laws of thermodynamics describe the transport of heat and work in macroscopic (\ie, large scale) processes and play a fundamental role in the physical sciences. The theory of thermodynamics is primarily an intellectual achievement of the $19^{th}$ century. The first analysis of heat engines was given by the French engineer Sadi Carnot in his seminal 1824 publication (`\textit{Reflections on the Motive Power of Fire and on Machines Fitted to Develop that Power}'~\cite{BibDB:Carnot,BibDB:Carnot2}), laying the foundations to the $2^{nd}$ law of thermodynamics. This paper marks the start of thermodynamics as a modern science~\cite{BibDB:HistoryOfThermo}.

The classical theory of thermodynamics was formulated in consistent form by giants like Joule, Clausius, Lord Kelvin, Gibbs and others. The atomic, or microscopic (\ie, small scale) approach to statistical thermodynamics was mainly developed through the pioneering work of Clausius, Maxwell, Boltzmann and Gibbs, laying the foundations to the more general discipline of statistical mechanics~\cite{BibDB:StatPhys}.

In particular, the $2^{nd}$  thermodynamic law, for quasi-static processes, linearly relates the change in the entropy, $dS$, to the amount of heat, $\dbar Q$, absorbed to a system at equilibrium, $\dbar Q=TdS$, where $T$ is the temperature of the system. However, the $2^{nd}$ law, in its classical formulation, is suited only for systems with energy Hamiltonian (function), $\mathcal{E}$, which is \emph{not} an explicit function of the temperature and fails to capture the physical behavior of systems with temperature-dependent energy levels. While such temperature-dependent energy function is uncommon in the study of natural and artificial systems in physics, it surprisingly does arise in modeling communication channels, like the popular Gaussian-noise communication channel~\cite{BibDB:BookCover,BibDB:BookProakis,BibDB:BookGallager2008}, as a (quasi-static) thermal system~\cite{BibDB:Sourlas,BibDB:Rujan,BibDB:BookNishimori}.

In this contribution, we generalize the $2^{nd}$ thermodynamic law to encompass systems with temperature-dependent Hamiltonian and obtain the generalized law \mbox{$\dbar Q=TdS+<d\mathcal{E}/dT>dT$}, where $<\cdot>$ denotes averaging over the standard Boltzmann distribution. Consequently, it allows for an alternative physical description to the Shannon-theoretic notions of information entropy and mutual information~\cite{BibDB:Shannon} via the thermodynamic  quantities of energy and temperature. As an example, the correct expressions for the mutual information of a Gaussian channel with Bernoulli-1/2 input and Gaussian input are computed from the thermodynamic representation, where the latter corresponds to the Shannon capacity.

Guo, Shamai and Verd\'{u}~\cite{BibDB:GSV} have recently revealed a simple, yet powerful relationship
(hereinafter termed GSV theorem) between information and
estimation theories. This cross-theory theorem bridges over the notions of
Shannon's mutual information and minimum mean-squre error (MMSE)
for the common additive white Gaussian noise channel.
Based on the thermodynamic expression of mutual information, the GSV theorem is naturally re-derived. This directly links the GSV theorem to the most profound laws of nature and gives the physical origin to this remarkable formula.


The paper is organized as follows. The thermodynamic background is discussed in Sections~\ref{sec_classical} and~\ref{sec_statistical}. In Section~\ref{sec_Gaussian} the Gaussian channel is represented through an equivalent thermal system giving a thermodynamic expression to the notions of information entropy and mutual information. In Section~\ref{sec_GSV} the GSV theorem is proven via thermodynamics. We conclude the paper in Section~\ref{sec_conclusion}.

We shall use the following notations. $P(\cdot)$ is used to
denote either a probability mass function (pmf), $\Pr(\cdot)$, or a
probability density function (pdf), $p(\cdot)$, depending on the random
variable having either discrete or continuous support, respectively.
Random variables are denoted by upper case letters and their
values denoted by lower case letters. The symbol $\mathbb{E}_{X}\{\cdot\}$ denotes expectation of the random object within the brackets with respect
to the subscript random variable. The natural logarithm, $\log$, is used.
The support of a variable $X$ is denoted by $\mathcal{X}$.

\section{Classical Thermodynamics}\label{sec_classical}
In this section we concisely summarize the fundamental results of classical
thermodynamics, essential to our discussion. The interested reader is
strongly encouraged to find a thorough introduction to thermodynamics in one
of numerous textbooks (\eg,~\cite{BibDB:Thermo}).

A thermodynamic system is defined as the part of the universe under
consideration, separated from the rest of the universe, referred to as
environment, surroundings or reservoir, by real or imaginary boundary. A
non-isolated thermodynamic system can exchange energy in the form of heat or
work with any other system. Heat is a process by which energy is added to a
system from a high-temperature source, or lost to a low-temperature sink. Work refers to forms of energy
transfer which can be accounted for in terms of changes in the macroscopic physical variables of the system (\eg, volume or pressure). For example, energy
which goes into expanding the volume of a system against an external
pressure, by driving a piston-head out of a cylinder against an external
force. Hereinafter we consider a thermal system which does not perform work,
mechanical or other. This thermal system is assumed to be in equilibrium,
that is all the descriptive macroscopic parameters of the system are
time-independent. We also assume that the process of exchanging heat is
infinitesimally quasi-static, \ie, it is carried out slowly enough that the
system remains arbitrarily close to equilibrium at all stages of the
process.

The underlying laws of thermodynamics consist of purely macroscopic
statements which make no reference to the microscopic properties of the system, \ie, to the molecules or particles of which they
consist. In the following statements we present the thermodynamic laws in a
form relevant to the thermal system under consideration.\footnote{For
conciseness the $0^{th}$ law's statement is omitted.}

\begin{LOT}
\begin{first}
   (conservation of energy) A system in equilibrium can be characterized by
a quantity $U$, called the `internal energy'. If the system is not isolated,
interact with another system and no work is done by it, the resulting change
in the internal energy can be written in the form
      \BE
      dU=\dbar Q,
      \EE where $\dbar Q$ is the amount of heat absorbed by the
system.\footnote{The  infinitesimal heat is denoted by $\dbar$ rather than
$d$ because, in mathematical terms, it is not an exact differential. The
integral of an inexact differential depends upon the particular path taken
through the space of (thermodynamic) parameters while the integral of an
exact differential depends only upon the initial and final states.
}
\end{first}
\begin{second} (definition of temperature) A system in equilibrium can be
characterized by a quantity $S$, called the `thermodynamic entropy'. If the
system is not isolated and undergoes a quasi-static infinitesimal process in
which it absorbs heat $\dbar Q$, then \BE\label{eq_second_law} dS=\frac{\dbar Q}{T},\EE where $T$
is a quantity characteristic of the system and is called the `absolute
temperature' of the system.
\end{second}
\begin{third} (zero entropy) The thermodynamic entropy $S$ of a system has a
limiting property that
      \BE
        \textrm{as } T\rightarrow 0_{+},\quad S\rightarrow S_{0}
      ,\EE where $S_{0}$ is a constant (usually $0$ in case of non-degenerate ground state energy) independent of all
parameters of the particular system.
\end{third}
\end{LOT}

Incorporating the three laws of thermodynamics together, a combined law describing the
thermodynamic entropy as an integration function over the temperature is obtained
\BE\label{eq_S}
S(T)=\int_{0}^{T}\frac{1}{\gamma}dU(\gamma)=\int_{0}^{T}\frac{1}{\gamma}\fd{U(\gamma)}{\gamma}d\gamma=\int_{0}^{T}\frac{C_{V}(\gamma)}{\gamma}d\gamma,
\EE where \BE C_{V}(T)\triangleq\frac{dU(T)}{dT} \EE is known as the heat
capacity (at constant volume $V$). The heat capacity is a non-negative
temperature-dependent measurable quantity describing the amount of heat
required to change the system's temperature by an infinitesimal degree.

Let us now define the inverse temperature
\mbox{$\beta\triangleq(k_{B}T)^{-1}$}, where the constant $k_{B}$ is the
Boltzmann constant. In this contribution we arbitrarily set $k_{B}=1$, thus
\mbox{$\beta=1/T$}. Hence, the entropy integral~(\ref{eq_S}) can be
rewritten in terms of the inverse temperature as \BE\label{eq_S_beta}
S(\beta)=-\int_{\beta}^{\infty}\gamma C_{V}(\gamma)d\gamma,
\EE where \BE C_{V}(\beta)\triangleq\frac{dU(\beta)}{d\beta}=-T^{2}C_{V}(T).
\EE

\section{Statistical Thermodynamics}\label{sec_statistical}
Moving to the microscopic perspective of statistical
thermodynamics~\cite{BibDB:Thermo,BibDB:Huang}, the probability, $P(X=x)$, of finding
the system in any one particular microstate, $X=x$, of
energy level $\mathcal{E}(X=x)$ is dictated according to the canonical Boltzmann
distribution~\cite{BibDB:StatPhys} \BE\label{eq_Boltzmann}
    P(X=x)=\frac{1}{\mathcal{Z}}\exp{\big(-\beta \mathcal{E}(X=x)\big)},
\EE where \BE
\mathcal{Z}\triangleq{\sum_{x\in\mathcal{X}}\exp{\big(-\beta
    \mathcal{E}(X=x)\big)}}
\EE is the partition (normalization) function, and the sum extends over all possible microstates of the system.

Applying the Boltzmann distribution, the macroscopic quantities of internal energy and entropy can be described microscopically as the average energy and the average $\{-\log P(X)\}$, respectively. Explicitly,
\BEA
U&=&\mathbb{E}_{X}\{\mathcal{E}(X)\}\label{eq_U},\\
S&=&\mathbb{E}_{X}\{-\log P(X)\},
\EEA and it can be easily verified that the following relation holds
\BE\label{eq_identity}
\log\mathcal{Z}=-\beta U+S.
\EE

\section{The Gaussian Channel As A Thermal System}\label{sec_Gaussian}
Consider a real-valued channel with input and output random variables $X$
and
$Y$, respectively, of the form \BE\label{eq_channel}
    Y=X+N,
\EE where \mbox{$N\sim\mathcal{N}(0,1/\snr)$} is a Gaussian noise
independent
of $X$, and $\snr\geq0$ is the channel's signal-to-noise ratio (SNR). The
input is taken from a probability distribution $P(X)$ that satisfies
$\mathbb{E}_{X}\{X^{2}\}<\infty$.\footnote{Similarly to the formulation in~\cite{BibDB:GSV} for $\mathbb{E}_{X}\{X^{2}\}=1$, $\snr$ follows
the usual notion of SNR. For $\mathbb{E}_{X}\{X^{2}\}\neq1$, $\snr$ can be
regarded as the gain in the output SNR due to the
channel.}

The Gaussian channel~(\ref{eq_channel}) (and any other communication channel) can be also viewed as a physical
system, operating under
the laws of thermodynamics. The microstates of the thermal system are
equivalent to the hidden values of the channel's input $X$. A comparison of
the channel's a-posteriori probability
distribution, given by \BEA
P{(X=x|Y=y)}&=&\frac{P(X=x)p(Y=y|X=x)}{p(Y=y)}\\&=&\frac{\sqrt{\snr}P(X=x)}{\sqrt{2\pi}p(Y=y)}\exp{\Big(-\frac{\snr}{2
}(y-x)^{2}\Big)}\\&=&\frac{\exp{\big(-\snr(-xy+\frac{x^{2}}{2}-\frac{\log
P(X=x)}{\snr})\big)}}{\sum_{x\in\mathcal{X}}\exp{\big(-\snr(-xy+\frac{x^{2}}{2}-\frac{\log
P(X=x)}{\snr})\big)}}\label{eq_denom},
\EEA with the Boltzmann distribution law~(\ref{eq_Boltzmann}) yields the
following
mapping of the inverse temperature and energy of the equivalent thermal
system \BEA\label{eq_snr_map}
    \snr&\rightarrow&\beta,\\
    -xy+\frac{x^{2}}{2}-\frac{\log P(X=x)}{\beta}&\rightarrow&
\mathcal{E}(X=x|Y=y;\beta).\label{eq_energy_map}
\EEA

Note that the temperature (\ie, the noise variance according to the
mapping~(\ref{eq_snr_map})) can be increased gradually from the absolute zero to its
target value $T=1/\beta$ in infinitesimally small steps, thus the
Gaussian channel system can remain arbitrarily close to equilibrium at all
stages of this process. Hence, the equivalent thermal system exhibits a
quasi-static infinitesimal process. Interestingly, the notion of
quasi-statics is reminiscent to the concept of Gaussian pipe in the SNR-incremental
Gaussian channel approach taken by Guo~\etal~\cite[Section \mbox{II-C}]{BibDB:GSV}.
Thus, we can apply the entropy integral~(\ref{eq_S_beta}) obtained from thermodynamics to the thermal system being equivalent to the Gaussian channel, yielding \BEA\label{eq_S_final}
S(\beta)=S(X|Y=y;\beta)=-\int_{\beta}^{\infty}\gamma
C_{V}(\gamma)d\gamma
=-\int_{\beta}^{\infty}\gamma\frac{dU(Y=y;\gamma)}{d\gamma}d\gamma,
\EEA where following~(\ref{eq_U}) the internal energy, $U(Y=y;\beta)=\mathbb{E}_{X|Y}\{\mathcal{E}(X|Y=y)\}$,
is the energy averaged over all possible values of $X$, given $y$.

The posterior information (Shannon) entropy, $H(X|Y;\beta)$, (in nats) of the channel
can be expressed via the thermodynamic entropy conditioned on $Y=y$, $S(X|Y=y;\beta)$~(\ref{eq_S_final}), as
\BEA\label{eq_S_XgivenY}
H(X|Y;\beta)=\mathbb{E}_{Y}\{S(X|Y=y;\beta)\}=-\mathbb{E}_{Y}\Bigg\{\int_{\beta}^{\infty}\gamma\frac{dU(Y;\gamma)}{d\gamma}d\gamma\Bigg\}. \EEA
The input's entropy can also be reformulated in a similar manner, since $H(X)=H(X|Y;\beta=0$). Hence,
\BEA
H(X)&=&-\mathbb{E}_{Y}\Bigg\{\int_{0}^{\infty}\gamma\frac{dU(Y;\gamma)}{d\gamma}d\gamma\Bigg\}.
\EEA

Now, the input-output mutual information can be described via thermodynamic
quantities, namely the energy, $\mathcal{E}$, and inverse temperature, $\beta$, as \BEA\label{eq_I}
I(X;Y)&=&I(\beta)\triangleq H(X)-H(X|Y;\beta)\\&=&-\mathbb{E}_{Y}\Bigg\{\int_{0}^{\beta}\gamma\frac{dU(Y;\gamma)}{d\gamma}d\gamma\Bigg\}\\
&=&-\big[\gamma\mathbb{E}_{Y}\{U(Y;\gamma)\}\big]_{0}^{\beta}+\mathbb{E}_{Y}\Bigg\{\int_{0}^{\beta}U(Y;\gamma)d\gamma\Bigg\}\label{eq_rhs}, \label{eq_MI_integral}\EEA
where~(\ref{eq_rhs}) is obtained using integration by parts.

Note that this thermodynamic interpretation to the mutual information holds
not only for the Gaussian channel, but also for any channel which can be
described by a thermal system exhibiting quasi-static heat transfer. In the following we illustrate the
utilization of~(\ref{eq_rhs}) by re-deriving the correct expression for the mutual information of a
Gaussian channel with Bernoulli-1/2 input.\\\,

\textbf{\emph{Example: Gaussian Channel with Bernoulli-$1/2$ Input}}

Since the input $X$ in this case is binary and equiprobable, \ie\, $P(X=1)=P(X=-1)=1/2$, the $X^{2}/2=1/2$ and $\log P(X=x)/\beta$ terms of the
Gaussian channel's energy~(\ref{eq_energy_map}) are independent of $X$ and can be dropped\footnote{Cancelled out by the same terms from the denominator in~(\ref{eq_denom}).}, leaving us
with the expression, independent of $\beta$,
\BE
\mathcal{E}(X=x|Y=y)=-xy.
\EE The a-posteriori probability mass function gets the form \BE
\Pr(X=x|Y=y)=\frac{\exp{(x\beta y)}}{\exp{(\beta y)}+\exp{(-\beta y)}},\quad
x\in\pm1.
\EE Hence, the internal energy is \BE
U(Y=y;\beta)=\mathbb{E}_{X|Y}\{\mathcal{E}(X|Y=y)\}=-y\tanh(\beta y). \EE The marginal pdf
of the output is then given by \BE
p(Y=y)=\frac{\sqrt{\beta}}{2\sqrt{2\pi}}\Bigg(\exp{\Big(-\frac{\beta}{2}(y-1)^{2}\Big)}+\exp{\Big(-\frac{\beta}{2}(y+1)^{2}\Big)}\Bigg).
\EE Thus, \BEA
-\beta\mathbb{E}_{Y}\{U(Y;\beta)\}&=&-\int_{-\infty}^{\infty}y\tanh
(\beta y)p(Y=y)dy\\&=&-\frac{\beta}{2\sqrt{2\pi}}\int_{-\infty}^{\infty}y\Bigg(
\exp{\Big(-\frac{\beta}{2}(y-1)^{2}\Big)}-\exp{\Big(-\frac{\beta}{2}(y+1)^{2}\Big)}\Bigg)dy\\&=&\frac{\beta}{2}(1-(-1))=\beta
\EEA and \BEA
\mathbb{E}_{Y}\Bigg\{\int_{0}^{\beta}U(Y;\gamma)d\gamma\Bigg\}&=&\int_{-\infty}^{\infty}p(Y=
y)\int_{0}^{\beta}U(Y=y;\gamma)d\gamma
dy\\&=&-\frac{1}{\sqrt{2\pi}}\int_{-\infty}^{\infty}\exp{\bigg(-\frac{y^{2
}}{2}\bigg)}\log\cosh{(\beta-\sqrt{\beta}y)}dy
\EEA giving, based on~(\ref{eq_rhs}), \BE
I(\beta)=\beta-\frac{1}{\sqrt{2\pi}}\int_{-\infty}^{\infty}\exp{\bigg(-\frac{y^{2
}}{2}\bigg)}\log\cosh{(\beta-\sqrt{\beta}y)}dy,
\EE which is identical to the known Shannon-theoretic result (see, \eg,~\cite[eq.
(18)]{BibDB:GSV} and~\cite[p. 274]{BibDB:BookBlahut}). \qed

\subsection{Generalized $2^{nd}$ Law}
When trying to repeat this exercise for a Gaussian input one finds
that~(\ref{eq_MI_integral}) fails to reproduce $I(\snr)=1/2\log(1+\snr)$,
which is the celebrated formula for the Shannon capacity of the Gaussian channel. This observation
can be explained as follows. The $2^{nd}$ thermodynamic law, as stated
above, holds only for systems with an energy function, $\mathcal{E}$, which is \emph{not} a
function of the temperature. While the temperature-independence of $\mathcal{E}$ is a
well-known conception in the investigation of both natural and artificial systems in
thermodynamics, and physics at large, such independence does not necessarily hold for communication channels and particularly for the Gaussian channel.
Actually, one can easily observe that the Gaussian channel's
energy~(\ref{eq_energy_map}) is indeed an explicit function of the
temperature via the $\log P(X=x)/\beta$ term, unless this term can be dropped by absorbing it into the partition function, as happens for equiprobable input distributions, like the discrete Bernoulli-1/2
or the continuous uniform distributions. This is exactly the reason why~(\ref{eq_MI_integral})
does succeed to compute correctly the mutual information for the particular case of a Bernoulli-$1/2$ input source.

In order to capture the temperature-dependent nature of the energy in systems like the communication channel, we generalize the formulation of the $2^{nd}$ law of
thermodynamics $dS=\dbar Q/T$.
\begin{second_G} (redefinition of temperature) If the thermal system is not isolated and undergoes a
quasi-static infinitesimal process in which it absorbs heat $\dbar Q$, then
\BE\label{eq_G_second_law} dS=\frac{\dbar Q}{T}-\frac{1}{T}\mathbb{E}_{X}\bigg\{\fd{\mathcal{E}(X)}{T}\bigg\}dT.\EE
\end{second_G}
\begin{proof}
The differential of the partition function's logarithm, $\log\mathcal{Z}$, can be written as
\BE
d\log\mathcal{Z}=\fd{\log\mathcal{Z}}{\beta}d\beta.
\EE Utilizing the identity~(\ref{eq_identity}), one obtains
\BE
dS=d(\beta U)+\fd{\log\mathcal{Z}}{\beta}d\beta.
\EE Since for $T$-dependent energy
\BE
\fd{\log\mathcal{Z}}{\beta}=-U-\beta\mathbb{E}_{X}\bigg\{\fd{\mathcal{E}(X)}{\beta}\bigg\},
\EE we get \BE
dS=d(\beta U)-Ud\beta-\beta\mathbb{E}_{X}\bigg\{\fd{\mathcal{E}(X)}{\beta}\bigg\}d\beta=\beta dU-\beta\mathbb{E}_{X}\bigg\{\fd{\mathcal{E}(X)}{\beta}\bigg\}d\beta,
\EE where the r.h.s. results from Leibniz's law (product rule). Recalling that according to the $1^{st}$ law $dU=\dbar Q$ concludes the proof.
\end{proof}

The generalized $2^{nd}$ law of thermodynamics~(\ref{eq_G_second_law}) has a clear physical interpretation. For simplicity, let us assume that the examined system is characterized by a comb of discrete energy levels $\mathcal{E}1,\mathcal{E}2,\ldots$ . The heat absorbed into the $T$-dependent system has the following dual effect: A first contribution of the heat, $dU-\mathbb{E}_{X}\{d\mathcal{E}(X)/dT\}dT$, increases the temperature of the system while the second contribution, $\mathbb{E}_{X}\{d\mathcal{E}(X)/dT\}dT$, goes for shifting the energy comb.  However, the shift of the energy comb does \emph{not} affect the entropy, since the occupation of each energy level remains the same, and the entropy is independent of the energy values which stand behind the labels $\mathcal{E}1,\mathcal{E}2,\ldots$. The change in the entropy can be done only by moving part of the occupation of one tooth of the energy comb to the neighboring teeth, and this can be achieved only by changing the temperature. Hence, the effective heat contributing to the entropy is $\dbar Q-\mathbb{E}_{X}\{d\mathcal{E}(X)/dT\}dT$, and this is the physical explanation to the generalized $2^{nd}$ law~(\ref{eq_G_second_law}). Note that for $T$-independent energy, the classical $2^{nd}$ law~(\ref{eq_second_law}) is immediately obtained.

Note also, that the $1^{st}$ law remains unaffected, $dU=\dbar Q$, since both ways of heat flow absorption into the system are eventually contributing to the average internal energy $U$. The generalized $2^{nd}$ law specifies the trajectory, the weight of each one of the two possible heat flows, at a given temperature.

The temperature, originally defined by the $2^{nd}$ law as $T=\dbar Q/dS$, is redefined now as\BE
\frac{1}{T}=\frac{dS/\dbar Q}{1-\frac{\mathbb{E}_{X}\{d\mathcal{E}(X)/dT\}}{\dbar Q/dT}}=\frac{dS/\dbar Q}{1-\frac{\mathbb{E}_{X}\{d\mathcal{E}(X)/dT\}}{C_{V}(T)}}.
\EE This redefinition has a more complex form and involves an implicit function of $T$ since the temperature appears on both sides of the equation.

Based on the generalized $2^{nd}$ law, the thermodynamic expression for the
mutual information~(\ref{eq_I}) of quasi-static (\eg, Gaussian) communication channels can be reformulated as  \BEA\label{eq_I2}
I(X;Y)&=&-\mathbb{E}_{Y}\Bigg\{\int_{0}^{\beta}\gamma\Big(dU(Y;\gamma)+\mathbb{E}_{X|Y}\bigg\{\fd{\mathcal{E}(
X|Y;\gamma)}{\gamma}\bigg\}d\gamma\Big)\Bigg\}\\&=&-\mathbb{E}_{Y}\Bigg\{\int_{0}^{\beta}\gamma\Big(\fd{U(Y;\gamma)}{\gamma}+\mathbb{E}_{X|Y}\bigg\{\fd{\mathcal{E}(
X|Y;\gamma)}{\gamma}\bigg\}\Big)d\gamma\Bigg\}\label{eq_I_beta}\\
&=&-\big[\gamma\mathbb{E}_{Y}\{U(Y;\gamma)\}\big]_{0}^{\beta}
+\mathbb{E}_{Y}\Bigg\{\int_{0}^{\beta}\Big(U(Y;\gamma)+\gamma\mathbb{E}_{X|Y}\bigg\{\fd{\mathcal{E}(X|Y;\gamma)}{\gamma}\bigg\}\Big)d\gamma\Bigg\}.
\label{eq_MI_integral2}\EEA
Again, the example of a Gaussian channel with, this time, standard Gaussian input is used to illustrate the
utilization of~(\ref{eq_MI_integral2}) for the correct derivation of the mutual information.\\\,

\textbf{\emph{Example: Gaussian Channel with $\mathcal{N}(0,1)$ Input}}

Since in this case $\log(P(X))/\beta=-x^{2}/(2\beta)$, the energy~(\ref{eq_energy_map})
of the Gaussian channel system becomes an explicit function of $\beta$,
given by
\BE
\mathcal{E}(X=x|Y=y;\beta)=-xy+\frac{x^{2}}{2}\bigg(\frac{1+\beta}{\beta}\bigg),
\EE and the derivative of this function with respect to $\beta$ yields \BE
\pd{\mathcal{E}(X=x|Y=y;\beta)}{\beta}=-\frac{x^{2}}{2\beta^{2}}. \EE The a-posteriori
probability density function is
\BE
p(X=x|Y=y;\beta)=\mathcal{N}\bigg(\frac{\beta
y}{1+\beta},\frac{1}{1+\beta}\bigg).
\EE Hence, the internal energy is \BE
U(Y=y;\beta)=\mathbb{E}_{X|Y}\{\mathcal{E}(X|Y=y;\beta)\}=-\frac{y^{2}\beta}{2(1+\beta)}+\frac{1}{2\beta}, \EE and the derivative of the energy averaged over all
possible inputs is \BE
\mathbb{E}_{X|Y}\Bigg\{\pd{\mathcal{E}(X|Y=y;\beta)}{\beta}\Bigg\}=-\frac{1}{2\beta^{2}}\bigg(\frac{1}{1+\beta}+\frac{\beta^{2}y^{2}}{(1+\beta)^{2}}\bigg).
\EE
The marginal pdf of the output is given by \BE
    p(Y=y)=\mathcal{N}\Bigg(0,\frac{1+\beta}{\beta}\Bigg).
\EE Thus, \BEA
    &&-\big[\gamma\mathbb{E}_{Y}\{U(Y;\gamma)\}\big]_{0}^{\beta}=\frac{\beta}{2}-\frac{1}{2}-\Big(-\frac{1}{2}\Big)=\frac{\beta}{2},
\EEA and \BEA
&&\mathbb{E}_{Y}\Bigg\{\int_{0}^{\beta}\Big(U(Y;\gamma)+\gamma\mathbb{E}_{X|Y}\bigg\{\fd{\mathcal{E}(X|Y;\gamma)}{\gamma}\bigg\}\Big)d\gamma\Bigg\} \\&&=\mathbb{E}_{Y}\Bigg\{\bigg[-\frac{1+\gamma}{2}+\frac{1+\gamma}{2\gamma}\log(1+\gamma)+\frac{1}{2}\log(1+\gamma)-\frac{Y^{2}}{2}\log(1+\gamma)-\frac{Y^{2}}{2(1+\gamma)}\bigg]_{0}^{\beta}\Bigg\}\\&&=
\mathbb{E}_{Y}\Bigg\{-\frac{\beta}{2}+\frac{1+\beta }{2\beta}\log(1+\beta)-\frac{1}{2}+\frac{1}{2}\log(1+\beta)-\frac{Y^{2}}{2}\log(1+\beta)-\frac{Y^{2}}{2(1+\beta)}+\frac{Y^{2}}{2}\Bigg\}\nonumber\\\\
&&=-\frac{\beta}{2}+\frac{1}{2}\log{(1+\beta)}
\EEA giving, based on~(\ref{eq_MI_integral2}), \BE
I(X;Y)=\frac{1}{2}\log{(1+\beta)}
\EE and the Shannon capacity~\cite{BibDB:Shannon} is derived from the perspective of thermodynamics.

\section{Guo-Shamai-Verd\'{u} Theorem}\label{sec_GSV}
In this section we prove the Guo-Shamai-Verd\'{u} (GSV) theorem from the $2^{nd}$ law
of thermodynamics for systems with $T$-dependent energy and the resulting thermodynamic representation of the mutual information. Thus, we show that this fascinating relation between
information theory and estimation theory is actually an evolution of the most
profound laws of nature.

To start with, let us restate the GSV theorem.
Consider a Gaussian channel of the form \BE\label{eq_channel_GSV}
    Y=\sqrt\snr X+N,
\EE where
\mbox{$N\sim\mathcal{N}(0,1)$} is a standard Gaussian noise independent
of $X$. The mutual information, $I(\snr)$~(\ref{eq_I}), and the minimum mean-square error, defined as \BEA \mmse(X|\sqrt\snr
X+N)&=&\mmse(\snr)=\mathbb{E}_{X,Y}\{(X-\mathbb{E}_{X|Y}\{X|Y;\snr\})^2\}, \EEA are both a function of $\snr$ and maintain the following relation.
\begin{GSV}{\cite[Theorem 1]{BibDB:GSV}}
For every input distribution $P(X)$ that satisfies
$\mathbb{E}_{X}\{X^{2}\}<\infty$, \BE\label{eq_GSV1}
    \fd{}{\snr}I(X;\sqrt{\snr}X+N)=\frac{1}{2}\mmse(X|\sqrt{\snr}X+N).
    \EE
\end{GSV}

Note that the GSV-based expression of the mutual information, \BE\label{eq_I_GSV}
I(X;\sqrt{\snr}X+N)=\frac{1}{2}\int_{0}^{\snr}\mmse(X|\sqrt{\gamma}X+N)d\gamma,
\EE resembles its thermodynamic expression~(\ref{eq_I_beta}) in the sense that both are an outcome of integration with respect to SNR (or inverse temperature, $\beta$). Hence, the roots of this integration in the GSV theorem may be attributed to the $2^{nd}$ law of thermodynamics. Note however to the opposite order of integration in the two expressions, where in the GSV expression~(\ref{eq_I_GSV}) the inner integration (within the definition of MMSE) is over $y$ and the outer integration is over SNR, and vice versa for the thermodynamic expression~(\ref{eq_I_beta}). Exchanging the order of integration in the latter (which is not trivial since $p(Y=y)$ is itself a function of $\beta$) yields, as we shall see, an integrand of $d\beta$ which is equal to $\mmse(\beta)/2$.
In the following proof Lemma~$1$ from~\cite{BibDB:GSV}, which underlines the main proof of the GSV theorem in~\cite{BibDB:GSV}, is proven directly from the thermodynamic description of the mutual information~(\ref{eq_I2}).

\begin{proof}
Adopting the SNR-incremental channel approach~\cite[Eq. (30)-(41)]{BibDB:GSV} and mapping again $\snr\rightarrow\beta$
\BE
\fd{I(X;Y)}{\beta}=\frac{I(X;Y_{1})-I(X;Y_{2})}{\delta}=\frac{I(\beta+\delta)-I(\beta)}{\delta}=\frac{I(X;Y_{1}|Y_{2
})}{\delta},
\EE where $I(X;Y_{1}|Y_{2})$ is the mutual information of the incremental
Gaussian channel \BE\label{eq_inc_channel}
Y_{1}=\sqrt{\delta}X+N,
\EE where $N$ is a standard Gaussian noise, $X$ is taken from the conditional probability $P(X|Y_{2})$,
$\delta\rightarrow0$ and \BE\label{eq_Y2} Y_{2}=X+\mathcal{N}(0,1/\beta) \EE with $X\sim P(X)$. Hence, we have to prove
\BE
\frac{I(X;Y_{1}|Y_{2})}{\delta}=\frac{1}{2}\mmse(\beta).
\EE Now, the principles of thermodynamics come into action. For this incremental channel~(\ref{eq_inc_channel}), the energy and its derivative are given by
\BEA
\mathcal{E}(X=x|Y_{1}=y_{1},Y_{2}=y_{2};\delta,\beta)&=&-\frac{xy_{1}}{\sqrt{\delta}}+\frac{x^{
2}}{2}-\frac{\log{P(X=x|Y_{2}=y_{2};\beta)}}{\delta}\EEA and \BEA
&&\fd{}{\delta}\mathcal{E}(X=x|Y_{1}=y_{1},Y_{2}=y_{2};\delta,\beta)=\frac{xy_{1}\delta^{-\frac{3}{2}}}{2}+\frac{\log{P(X=x|Y_{2}=y_{2};\beta)}}{\delta^{2}}.
\EEA
Using the thermodynamic expression for the mutual information~(\ref{eq_I2}) and recalling
that $\delta\rightarrow0$, one gets \BEA
I(X;Y_{1}|Y_{2}=y_{2})&=&-\mathbb{E}_{Y_{1}|Y_{2}}\Bigg\{\int_{0}^{\delta}\gamma
\Big(dU(Y_{1}|Y_{2}=y_{2};\gamma,\beta)+\mathbb{E}_{X|Y_{1},Y_{2}}\bigg\{\pd{}{\gamma}{\mathcal{E}(X|Y_{1},Y_{2}=y_{2};\gamma,\beta)}\bigg\}d\gamma\Big)\Bigg\}
\nonumber\\\\&=&-\mathbb{E}_{Y_{1}|Y_{2}}\Bigg\{
\mathbb{E}_{X|Y_{1},Y_{2}}\bigg\{\delta \mathcal{E}(X|Y_{1},Y_{2}=y_{2};\delta,\beta)+\delta^{2}\fd{}{\delta}{\mathcal{E}(X|Y_{1},Y
_{2}=y_{2};\delta,\beta)}\bigg\}\Bigg\}
\\&=&-\mathbb{E}_{Y_{1}|Y_{2}}\Bigg\{-\frac{\mathbb{E}_{X|Y_{1},Y_{2}}\{X|Y_{1},Y_{2}=y_{2}\}Y_{1}\sqrt{\delta}}{2}+\frac{\delta
\mathbb{E}_{X|Y_{1},Y_{2}}\{X^{2}|Y_{1},Y_{2}=y_{2}\}}{2}\Bigg\}\\&=&
-\mathbb{E}_{Y_{1}|Y_{2}}\Bigg\{-\frac{\sqrt{\delta}}{2}Y_{1}\mathbb{E}_{X|Y_{2}}\{X|Y_{2}=y_{2}\}+\frac{\delta}{2}\mathbb{E}_{X|Y_{2}}\{{X^{2}|Y_{2}=y_{2}}\}\Bigg\}
\label{eq_move1}\\&=&\frac{\delta}{2}\Big(-\mathbb{E}_{X|Y_{2}}^{2}\{X|Y_{2}=y_{2}\}+\mathbb{
E}_{X|Y_{2}}\{X^{2}|Y_{2}=y_{2}\}\Big)\label{eq_move2}
\\&=&\frac{\delta}{2}\mathbb{E}_{X|Y_{2}}\{(X-\mathbb{E}_{X|Y_{2}}\{X|Y_{2}=y_{2}\})^{2}|Y_{2}=y_{2}
\},\EEA where~(\ref{eq_move1}) is based on the fact that for an infinitesimal $\snr$, $\delta\rightarrow0$, the expectation \mbox{$\mathbb{E}_{X|Y_{1},Y_{2}}\rightarrow\mathbb{E}_{X|Y_{2}}$}, while~(\ref{eq_move2}) results from the independence of $N$~(\ref{eq_inc_channel}) with $Y_{2}$~(\ref{eq_Y2})~\cite[Section II-C]{BibDB:GSV}.
Averaging over $Y_{2}$ on both sides of the equation, we obtain the desired result
\BE
I(X;Y_{1}|Y_{2})=\frac{\delta}{2}\mathbb{E}_{X,Y_{2}}\{(X-\mathbb{E}_{X|Y_{2}}\{X|Y_{2}\})^{2}\}
=\frac{\delta}{2}\mmse(\beta).
\EE
\end{proof}

\section{conclusion}\label{sec_conclusion}
In this paper, the mutual information of Gaussian channels is described via thermodynamic terminology. As a byproduct, an intimate link is revealed between the GSV theorem and the basic laws of thermodynamics.
More generally, the revised $2^{nd}$ law for thermal systems with temperature-dependent energy levels enables to quantitatively bridge between the realm of thermodynamics and information theory for quasi-static systems. It is anticipated that this substantial theoretical connection between the foundation laws of thermodynamics and information theory will open a horizon for new discoveries and development in the study of both artificial and natural systems, towards a possibly more synergetic foundation to these two vital disciplines.


\end{document}